\documentclass[aps,prb,twocolumn,showpacs,amsmath,amssymb]{revtex4-1}
\usepackage{graphicx}
\usepackage{dcolumn}
\usepackage{bm}

\begin{document}


\title{Symmetry-mode-based classical and quantum mechanical formalism of lattice dynamics}

\author{Tsezar F. Seman$^1$, Jichan Moon$^2$ and K. H. Ahn$^{1}$}\email{kenahn@njit.edu}
\affiliation{$^1$Department of Physics, New Jersey Institute of Technology, Newark, New Jersey 07102, USA
\\
$^2$Department of Physics, Konkuk University, Seoul, 143-701, South Korea}


\begin{abstract}
    We present classical and quantum mechanical descriptions of
    lattice dynamics, from the atomic to the continuum scale, using atomic scale symmetry modes and their
    constraint equations. This approach is demonstrated for a one-dimensional
    chain and a two-dimensional square lattice with a monatomic basis.
    For the classical description, we find that rigid modes, in
    addition to the distortional modes found before, are necessary
    to describe the kinetic energy.
    The long wavelength limit of the kinetic energy terms expressed in terms of atomic scale modes
    is shown to be consistent with the continuum theory, and the leading
    order corrections are obtained.
    For the quantum mechanical description, we find conjugate momenta
    for the atomic scale symmetry modes.
    In direct space, graphical rules for their commutation relations
    are obtained.  Commutation
    relations in the reciprocal space are also calculated. As an example, phonon modes
    are analyzed in terms of symmetry modes.
    We emphasize that the approach based
    on atomic scale symmetry modes could be useful, for example,
    for the description of multiscale lattice dynamics and
    the dynamics near structural phase transition.
\end{abstract}


\maketitle
\section{Introduction}
The dynamics at nanometer length scale has been a focus of recent attention.~\cite{Rini07}
In particular, materials with competing ground states, such as high temperature superconducting
cuprates and colossal magnetoresistive manganites,~\cite{Jin94, Millis98, Salamon01}
often show dynamic nanometer scale features,
for example, stripes in cuprates~\cite{Tranquada95, Kivelson03} and anisotropic correlations in manganites.~\cite{Kiryukhin04, Ahn04}
Furthermore, recent advances in time-resolved x-ray technique
have allowed experimenters to directly probe lattice dynamics in atomic scale.~\cite{Gaffney05}
It is believed that understanding these nanoscale features and their dynamics
is essential to explain macroscopic properties of these materials.

For the description of mesoscopic scale domain structures and their dynamics,
phenomenological Ginzburg-Landau formalism has been very successful.~\cite{Shenoy99, Lookman03}
One of the keys for such a success
is the use of symmetry in the definition of variables, which makes the selection of
free energy terms self-evident.
Motivated by the success of the Ginzburg-Landau approach for the continuum,
symmetry-based atomic scale description of lattice distortions has been recently proposed,
and demonstrated for a two-dimensional square lattice.~\cite{Ahn03}
In this approach, atomic scale symmetry modes are defined on a plaquette of atoms,
and are used to express potential energy terms associated with lattice distortions.
This method has been used to understand atomic scale structure of twin boundaries,~\cite{Ahn03}
antiphase boundaries and their electronic textures,~\cite{Ahn05}
strain-induced metal-insulator phase coexistence in manganites,~\cite{Ahn04}
superconducting order parameter textures around structural defects,~\cite{Zhu03}
and the coupling between electronic nematic order parameter and structural domains
in metamagnets near a quantum critical point.~\cite{Doh07}
Thus far, this approach has been used for static lattices, or the relaxation of lattice distortions
introduced through the Euler method,~\cite{Shenoy99} which does not require kinetic energy terms.
In the current paper, we present our study on how the approach based on atomic scale
symmetry modes can be extended to describe lattice dynamics,
within the scope of both classical and quantum mechanical formalism.
In Section II, we discuss how to express kinetic energy term in symmetry modes,
present our study within the formalism of classical mechanics and compare our result
with the continuum results.~\cite{Lookman03}
We formulate quantum mechanics
in terms of atomic scale symmetry modes in Section III, with
a summary given in Section IV.
Appendix A presents a simple demonstration of symmetry-mode-based approach for lattice dynamics,
that is, the phonon mode analysis in terms of symmetry modes.

\section{Classical formalism}
\subsection{One-dimensional lattice with a monatomic basis}
\begin{figure}[h]
        \includegraphics[scale=1.0, clip=true]{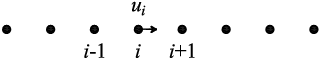}
        \caption{One-dimensional lattice with a monatomic basis.}
        \label{fig:chain}
    \end{figure}

\begin{figure}[h]
        \includegraphics[scale=1.0, clip=true]{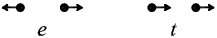}
        \caption{Symmetry modes for the one-dimensional chain in Fig.~\ref{fig:chain}.}
        \label{fig:linearmode}
    \end{figure}

Using a one-dimensional lattice with a monatomic basis shown in Fig.~\ref{fig:chain},
we demonstrate the concept of the mode-based description of lattice dynamics.
The displacements of atoms are represented by $u_i$, where $i$ being an index for the site.
To be specific, we assume that the interaction between the nearest neighbor atoms
are described by a spring with a spring constant $K$ while other potential energy terms are negligible,
as represented by the following Lagrangian,
    \begin{equation}
        L_{\rm chain} = \sum_{i} \frac{1}{2} M \dot{u}_i^2 -
            \frac{1}{2} K(u_{i+1}-u_i)^2,
            \label{eq:L_chain_u}
    \end{equation}
where $M$ is the mass of the atom.
We take a two-atom unit as a motif for this lattice, and define the symmetry modes,
$e(i)$ and $t(i)$,
    \begin{eqnarray}
        e(i) &\equiv& \frac{1}{\sqrt{2}}(u_{i+1} - u_i), \label{eq:chain.e} \\
        t(i) &\equiv& \frac{1}{\sqrt{2}}(u_{i+1} + u_i), \label{eq:chain.t}
    \end{eqnarray}
    where a normalization factor is chosen according to the number of displacement
variables in the definition.
    These modes are also shown in Fig.~\ref{fig:linearmode}.
    \\
The two variables, $e$ and $t$,
correspond to the distortion and rigid
translation of the motif, respectively.
Since the two modes are defined from one physically independent displacement variable at each site $i$,
these modes are related through one constraint equation shown below in the reciprocal space and
direct space, respectively.
    \begin{equation} \label{eq:constraint.1d}
       f(k) \equiv (e^{ik} + 1)e(k) - (e^{ik} - 1)t(k) = 0,
    \end{equation}
    \begin{equation} \label{eq:constraint.1d.direct}
        e(i+1)+e(i) - t(i+1)+t(i)=0.
    \end{equation}
In terms of these modes, the Lagrangian in Eq.~(\ref{eq:L_chain_u}) is expressed
in the following way
    \begin{equation} \label{eq:kinetic.1d.real}
    L_{\rm chain}=\sum_{i} \frac{1}{2}\left(\frac{M}{2}\right) [\dot{e}(i)^2 + \dot{t}(i)^2 ]
        - \frac{1}{2} (2K) e(i)^2.
    \end{equation}
The result shows that introduction of atomic scale rigid modes, such as $t$, which are
not considered in Ref.~\onlinecite{Ahn03}, allows
kinetic energy term being expressed in a quadratic form.
To obtain equations of motion,
we modify $L_{\rm chain}$ with a Lagrange multiplier $\lambda(k)$,
as shown below.
   \begin{eqnarray}
        \tilde{L}_{\rm chain} &=& \sum_k \frac{1}{2}\left(\frac{M}{2}\right) [\dot{e}(k) \dot{e}(-k)
            + \dot{t}(k) \dot{t}(-k)]  \nonumber \\
        && - \frac{1}{2}(2K) e(k) e(-k) + \lambda(k) f(-k).
        \label{eq:lagrangian.1d}
    \end{eqnarray}
The Lagrangian formalism of dynamics leads to the two
equations of motion,
    \begin{eqnarray}
        \frac{M}{2} \ddot{e}(k) + 2K e(k) - \lambda(k) (e^{-ik} + 1) &=& 0, \\
        \frac{M}{2} \ddot{t}(k) + \lambda(k) (e^{-ik} - 1) &=& 0,
    \end{eqnarray}
and a well-known dispersion relation
for the one-dimensional chain,~\cite{Kittel}
 \begin{equation}
        \omega = \sqrt{\frac{K}{M}(1-\cos{k})}.
    \end{equation}
This result shows that the lattice dynamics can be studied within the framework
of atomic scale symmetry modes and their constraint equations, without using the displacement
variables explicitly.

Anharmonicity of one-dimensional chains is important to understand
non-linear excitations, such as
solitons, kink-solitons, intrinsically localized modes, and breathers.~\cite{Chen96,Kosevich08}
Atomic scale modes, $e$ and $t$, found here can be used to incorporate
such anharmonicity into the Hamiltonian, which, along with their constraint equations,
would provide a formalism to study the dynamics of non-linear excitations in one-dimensional chains.
In the next subsection, we demonstrate
how the mode-based approach is applied to lattice dynamics for
a two-dimensional square lattice with a monatomic basis.

\subsection{Two-dimensional square lattice with a monatomic basis}
\begin{figure}[b]
        \includegraphics[scale=1.0, clip=true]{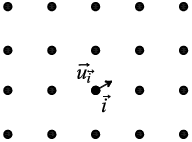}
        \caption{Two-dimensional square lattice with a monatomic basis.}
        \label{fig:squarelattice}
    \end{figure}

\begin{figure}[!hbp]
        \includegraphics[scale=1.0, clip=true]{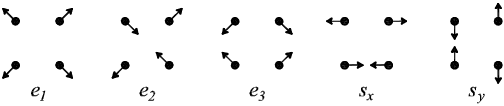}
        \caption{Distortional symmetry modes of the motif for the two-dimensional square lattice
        with a monatomic basis in Fig.~\ref{fig:squarelattice}.}
        \label{fig:squaremode}
    \end{figure}

Symmetry-based atomic scale description of lattice distortions
for a two-dimensional square lattice with a monatomic basis,
shown in Fig.~\ref{fig:squarelattice},
has been studied in Ref.~\onlinecite{Ahn03}, where
dilatational $e_1$, shear $e_2$, and deviatoric $e_3$ modes, and short wavelength modes, $s_x$ and $s_y$, are defined,
as shown in Fig.~\ref{fig:squaremode}.
In terms of displacement variables $u^x_{\vec{i}}$ and $u^y_{\vec{i}}$,
and $i_x$ and $i_y$ representing site indices, these distortional symmetry modes are expressed as follows.
    \begin{eqnarray}
        e_{1}(\vec{i})&=& \frac{1}{2\sqrt{2}} \left(-u^{x}_{\vec{i}}-u^{y}_{\vec{i}}+u^{x}_{\vec{i}+10}-u^{y}_{\vec{i}+10} \right. \nonumber \\
            && \left. -u^{x}_{\vec{i}+01}+u^{y}_{\vec{i}+01}+u^{x}_{\vec{i}+11}+u^{y}_{\vec{i}+11}\right), \label{eq:def-e1} \\
        e_{2}(\vec{i})&=& \frac{1}{2\sqrt{2}} \left(-u^{x}_{\vec{i}}-u^{y}_{\vec{i}}-u^{x}_{\vec{i}+10}+u^{y}_{\vec{i}+10} \right. \nonumber \\
            && \left. +u^{x}_{\vec{i}+01}-u^{y}_{\vec{i}+01}+u^{x}_{\vec{i}+11}+u^{y}_{\vec{i}+11}\right), \\
        e_{3}(\vec{i})&=& \frac{1}{2\sqrt{2}} \left(-u^{x}_{\vec{i}}+u^{y}_{\vec{i}}+u^{x}_{\vec{i}+10}+u^{y}_{\vec{i}+10} \right. \nonumber \\
            && \left. -u^{x}_{\vec{i}+01}-u^{y}_{\vec{i}+01}+u^{x}_{\vec{i}+11}-u^{y}_{\vec{i}+11}\right), \label{eq:def-e3} \\
        s_{x}(\vec{i})&=& \frac{1}{2} \left(u^{x}_{\vec{i}}-u^{x}_{\vec{i}+10}-u^{x}_{\vec{i}+01}+u^{x}_{\vec{i}+11}\right), \\
        s_{y}(\vec{i})&=& \frac{1}{2} \left(u^{y}_{\vec{i}}-u^{y}_{\vec{i}+10}-u^{y}_{\vec{i}+01}+u^{y}_{\vec{i}+11}\right). \label{eq:def-sy}
    \end{eqnarray}
Instead of $s_x$ and $s_y$ modes, the following $s_+$ and $s_-$ modes can be also used.
\begin{eqnarray}
s_{+}(\vec{i}) &=& \frac{1}{\sqrt{2}} [ s_{x}(\vec{i})+s_{y}(\vec{i})],  \\
s_{-}(\vec{i}) &=& \frac{1}{\sqrt{2}} [s_{x}(\vec{i})-s_{y}(\vec{i})].
\end{eqnarray}
These five modes have been used to express various harmonic and anharmonic potential
energy terms,~\cite{Ahn03, Ahn04, Ahn05}
but are not sufficient to represent kinetic energy terms in a simple quadratic form.

\begin{figure}[h]
        \includegraphics[scale=1.0, clip=true]{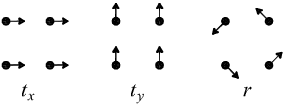}
        \caption{Rigid symmetry modes of the motif for the two-dimensional square lattice with a monatomic basis
        in Fig.~\ref{fig:squarelattice}.}
        \label{fig:rigid2d}
    \end{figure}

In current work, we show that additional modes
associated with the rigid motion of the motif, similar to the mode $t$ in the one-dimensional chain,
allow a formalism entirely based on symmetry modes without
resorting to displacement variables.
Three rigid modes for the two-dimensional
square lattice are shown in Fig.~\ref{fig:rigid2d} and are defined as follows.
    \begin{eqnarray}
        t_{x}(\vec{i})&=& \frac{1}{2} \left(u^{x}_{\vec{i}}+u^{x}_{\vec{i}+10}+u^{x}_{\vec{i}+01}+u^{x}_{\vec{i}+11}\right), \\
        t_{y}(\vec{i})&=& \frac{1}{2} \left(u^{y}_{\vec{i}}+u^{y}_{\vec{i}+10}+u^{y}_{\vec{i}+01}+u^{y}_{\vec{i}+11}\right), \\
        r(\vec{i})    &=& \frac{1}{2\sqrt{2}} \left(u^{x}_{\vec{i}}-u^{y}_{\vec{i}}+u^{x}_{\vec{i}+10}+u^{y}_{\vec{i}+10} \right. \nonumber \\
                      & & \left. -u^{x}_{\vec{i}+01}-u^{y}_{\vec{i}+01}-u^{x}_{\vec{i}+11}+u^{y}_{\vec{i}+11} \right). \label{eq:def-r}
    \end{eqnarray}  \\
    \\
The first two modes, $t_x$ and $t_y$, correspond to rigid translations of the motif
along $x$ and $y$ direction, and the mode $r$ represents a rigid rotation of the motif.
Following $t_+$ and $t_-$ modes can be also used as alternatives to $t_x$ and $t_y$.
   \begin{eqnarray}
t_{+}(\vec{i}) &=& \frac{1}{\sqrt{2}} [ t_{x}(\vec{i})+t_{y}(\vec{i})],  \\
t_{-}(\vec{i}) &=& \frac{1}{\sqrt{2}} [t_{x}(\vec{i})-t_{y}(\vec{i})].
\end{eqnarray}

Straight-forward expansion shows that the kinetic energy of the lattice
is expressed in terms of the eight symmetry modes in the following quadratic form,
with $M$ being the mass of the atom.
   \begin{eqnarray}
        T_{\rm sq} &=& \sum_{\vec{i}}\frac{1}{2}M[(\dot{u}^{x}_{\vec{i}})^2+
            (\dot{u}^{y}_{\vec{i}})^2] \label{eq:sq.lat.u} \\
        &=& \sum_{\vec{i}}\frac{1}{2}\left(\frac{M}{4}\right)
            [\dot{e}_{1}(\vec{i})^2+\dot{e}_{2}(\vec{i})^2+\dot{e}_{3}(\vec{i})^2
            +\dot{s}_{x}(\vec{i})^2 \nonumber \\
        && +\dot{s}_{y}(\vec{i})^2+\dot{t}_{x}(\vec{i})^2 +\dot{t}_{y}(\vec{i})^2
            +\dot{r}(\vec{i})^2].
        \label{eq:sq.lat.kinetic}
    \end{eqnarray}
As discussed in Ref.~\onlinecite{Ahn03}, constraint equations are found
from the relations between symmetry modes and displacement variables
in the reciprocal space.
We first represent ($u^x_{\vec{k}}$, $u^y_{\vec{k}}$) in terms of ($s_x(\vec{k})$, $s_y(\vec{k})$)
by inverting the linear relations between them,
and replace in the expressions with other modes,
which lead to six constraint equations.~\cite{footnote}
    \begin{eqnarray}
        & \sin{\frac{k_x}{2}}\cos{\frac{k_y}{2}}s_{x}(\vec{k}) +
            \cos{\frac{k_x}{2}}\sin{\frac{k_y}{2}}s_{y}(\vec{k})- \nonumber\\
        & - \sqrt{2} i \sin{\frac{k_x}{2}}\sin{\frac{k_y}{2}}e_{1}(\vec{k})
            = 0, \label{eq:constraint1.2d} \\
        & \cos{\frac{k_x}{2}}\sin{\frac{k_y}{2}}s_{x}(\vec{k}) +
            \sin{\frac{k_x}{2}}\cos{\frac{k_y}{2}}s_{y}(\vec{k})- \nonumber\\
        & - \sqrt{2} i \sin{\frac{k_x}{2}}\sin{\frac{k_y}{2}}e_{2}(\vec{k})
            = 0, \label{eq:constraint2.2d} \\
        & \sin{\frac{k_x}{2}}\cos{\frac{k_y}{2}}s_{x}(\vec{k}) -
            \cos{\frac{k_x}{2}}\sin{\frac{k_y}{2}}s_{y}(\vec{k})- \nonumber\\
        & - \sqrt{2} i \sin{\frac{k_x}{2}}\sin{\frac{k_y}{2}}e_{3}(\vec{k})
            = 0, \label{eq:constraint3.2d} \\
        & \cos{\frac{k_x}{2}}\sin{\frac{k_y}{2}}s_{x}(\vec{k}) -
            \sin{\frac{k_x}{2}}\cos{\frac{k_y}{2}}s_{y}(\vec{k})+ \nonumber\\
        & + \sqrt{2} i \sin{\frac{k_x}{2}}\sin{\frac{k_y}{2}}r(\vec{k})
            = 0, \label{eq:constraint4.2d} \\
        & \cos{\frac{k_x}{2}}\cos{\frac{k_y}{2}}s_{x}(\vec{k}) +
            \sin{\frac{k_x}{2}}\sin{\frac{k_y}{2}}t_{x}(\vec{k}) = 0, \label{eq:constraint5.2d} \\
        & \cos{\frac{k_x}{2}}\cos{\frac{k_y}{2}}s_{y}(\vec{k}) +
            \sin{\frac{k_x}{2}}\sin{\frac{k_y}{2}}t_{y}(\vec{k}) = 0. \label{eq:constraint6.2d}
    \end{eqnarray}
Modified Lagrangian for the square lattice is now
\begin{equation}
        \tilde{L}_{\rm sq}=T_{\rm sq}-V+\sum_{n=1}^6\sum_{\vec{k}}\lambda_{n}(\vec{k})f_{n}(-\vec{k}), \label{eq:Lagrangian}
    \end{equation}
where $V$ is the potential energy, $\lambda_{n}(\vec{k})$ are Lagrange multipliers,
and $f_{n}(\vec{k})=0$ are the six constraint equations, Eqs.~(\ref{eq:constraint1.2d}) - (\ref{eq:constraint6.2d}).
By solving the Lagrangian equations, we find dynamic properties of the
lattice in terms of the symmetry modes.
As with the Ginzburg-Landau approach being useful for the description of mesoscale dynamics,
we expect that the approach based on atomic scale symmetry modes would be useful
for the description of atomic scale dynamics, particularly,
when anharmonicity plays an essential role.

\subsection{Comparison with continuum description of lattice dynamics}

We compare the atomic scale theory developed in the previous subsection
with an existing continuum theory of lattice dynamics.
Either by using the definitions, Eqs.~(\ref{eq:def-e1})~and~(\ref{eq:def-e3}), or by using the constraint equations,
we express the kinetic energy for the square lattice, Eqs.~(\ref{eq:sq.lat.u}) and (\ref{eq:sq.lat.kinetic}),
in terms of $e_1$ and $e_3$,
    \begin{equation}
        T_{\rm sq} = \sum_{\vec{k}} \sum_{s=1,3} \sum_{s'=1,3}
        \frac{1}{2} M \gamma_{ss'}(\vec{k})
            \dot{e}_s(\vec{k}) \dot{e}_{s'}(-\vec{k}),
    \end{equation}
where
\begin{eqnarray}
        \gamma_{11}(\vec{k})=\gamma_{33}(\vec{k}) &=& \frac{1-\cos k_x \cos k_y}{\sin^2 k_x \sin^2 k_y} \label{gamma-ii}, \\
        \gamma_{13}(\vec{k})=\gamma_{31}(\vec{k}) &=& \frac{\cos k_x-\cos k_y}{\sin^2 k_x \sin^2 k_y} \label{gamma-ij}.
    \end{eqnarray}
To compare with a continuum theory,
we take the long wavelength limit, and obtain the following leading order
term for $\gamma_{ss'}$,
    \begin{equation}
        \gamma_{s s'}^{(0)}(\vec{k}) = \left(\begin{array}{c c}
              \displaystyle \frac{k_x^2+k_y^2}{2 k_x^2 k_y^2}
            & \displaystyle \frac{k_y^2-k_x^2}{2 k_x^2 k_y^2} \\ \\
              \displaystyle \frac{k_y^2-k_x^2}{2 k_x^2 k_y^2}
            & \displaystyle \frac{k_x^2+k_y^2}{2 k_x^2 k_y^2}
        \end{array}\right). \label{rho-0}
    \end{equation}
    This term is identical to Eq.~(3.12a) in Ref.~\onlinecite{Lookman03}
    ($e_3$ here corresponds to $e_2$ in Ref.~\onlinecite{Lookman03}),
which Lookman {\it et al.} have used as continuum kinetic energy
to study underdamped dynamics of strains in proper ferroelastic materials.
The next order term to the above continuum limit is
as follows.
    \begin{equation}
        \gamma_{s s'}^{(1)}(\vec{k}) = \left(\begin{array}{c c}
              \displaystyle \frac{1}{12}+\frac{k_x^4+k_y^4}{8 k_x^2 k_y^2}
            & \displaystyle \frac{k_y^4-k_x^4}{8 k_x^2 k_y^2} \\ \\
              \displaystyle \frac{k_y^4-k_x^4}{8 k_x^2 k_y^2}
            & \displaystyle \frac{1}{12}+\frac{k_x^4+k_y^4}{8 k_x^2 k_y^2}
        \end{array}\right). \label{rho-1}
    \end{equation}
This term, or better Eqs.~(\ref{gamma-ii}) and (\ref{gamma-ij}), can be used
to study the dynamics of proper ferroelastic materials
on the atomic scale.
    The following long wavelength limit of the atomic scale modes
    shows directly what they correspond to in the continuum theory.
   \begin{eqnarray}
        e_1(\vec{i}) &=& \frac{1}{\sqrt{2}}[\nabla_x u^x_{\vec{i}} + \nabla_y u^y_{\vec{i}}], \\
        e_2(\vec{i}) &=& \frac{1}{\sqrt{2}}[\nabla_x u^y_{\vec{i}} + \nabla_y u^x_{\vec{i}}], \\
        e_3(\vec{i}) &=& \frac{1}{\sqrt{2}}[\nabla_x u^x_{\vec{i}} - \nabla_y u^y_{\vec{i}}], \\
        r(\vec{i})   &=& \frac{1}{\sqrt{2}}[\nabla_x u^y_{\vec{i}} - \nabla_y u^x_{\vec{i}}], \\
        s_x(\vec{i}) &=& \frac{1}{2} \nabla_x\nabla_y u^x_{\vec{i}}, \\
        s_y(\vec{i}) &=& \frac{1}{2} \nabla_x\nabla_y u^y_{\vec{i}}, \\
        t_x(\vec{i}) &=& 2u^x_{\vec{i}}, \\
        t_y(\vec{i}) &=& 2u^y_{\vec{i}}.
    \end{eqnarray}
In $k \rightarrow 0$ limit, the correspondence of these modes to the displacements $u$ are
   \begin{eqnarray}
        t_x, t_y &\sim& u, \nonumber \\
        e_1, e_2, e_3, r &\sim& k u, \label{eq:klimit}\\
        s_x, s_y &\sim& k^2 u. \nonumber
    \end{eqnarray}
The comparison shows that our approach is a natural extension of the continuum theory to the atomic scale,
and is suitable for multiscale description of lattice dynamics.

\section{Quantum mechanical formalism}
\subsection{One-dimensional lattice with a monatomic basis}
It is necessary to consider quantum mechanical aspects of lattice dynamics
for phenomena such as
low temperature specific heat, electron-phonon interaction, and polarons.
In this section, we extend the symmetry-based atomic scale description of lattice dynamics
to the quantum mechanical formalism.
Commutation relations between the coordinate operators and their conjugate momentum operators
lie at the core of quantum mechanics, which we establish here for the symmetry modes.

First, we consider the one-dimensional chain studied in Section II.A.
The conjugate momenta for the two modes, $P_e(i)$ and $P_t(i)$
are
    \begin{eqnarray}
        P_e(i) &=& \frac{\partial L}{\partial{\dot{e}}(i)}
            = \frac{M}{2}\dot{e}(i) = \frac{1}{2\sqrt{2}}(p_{i+1} - p_{i}), \\
        P_t(i) &=& \frac{\partial L}{\partial{\dot{t}}(i)}
            = \frac{M}{2}\dot{t}(i) = \frac{1}{2\sqrt{2}}(p_{i+1} + p_{i}),
    \end{eqnarray}
where $p_i$ represents the momentum of the atom at site $i$.
From known commutation relations between momentum and displacement operators
$\hat{p}_i$ and $\hat{u}_j$,
we find the following commutation relations between the operators for modes and their conjugate momenta
with the same site index $i$,
    \begin{equation*}
        {[P_{a}(i), b(i)]}=\frac{\hbar}{2i} \delta_{ab},
    \end{equation*}
    where $a, b \in \{e, t\}$.
    The 1/2 factor is related to the number of atoms in each motif.
Unlike displacement variables,
the commutation relation between a mode at $i$ and a conjugate momentum
at $i+1$ or $i-1$ is not zero, since they share an atom, as shown below.
        \begin{equation*}
        {[P_{t}(i), t(i \pm 1)]}={[P_{e}(i), t(i+1)]} = {[P_{t}(i), e(i-1)]} =  \frac{\hbar}{4i},
        \end{equation*}
        \begin{equation*}
        {[P_{e}(i), e(i \pm 1)]}={[P_{t}(i), e(i+1)]} = {[P_{e}(i), t(i-1)]} = \frac{-\hbar}{4i}.
    \end{equation*}
    The commutation relations between the momentum and the mode, defined at sites
    further than the nearest neighbors, vanish.

\begin{figure}[h]
        \includegraphics[scale=1.0, clip=true]{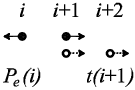}
        \caption{Graph to find commutation relation, $[P_e(i),t(i+1)]$.}
        \label{fig:commutationex1.1d.graph}
    \end{figure}

    The above relations are also established graphically.
    For example,  ${[P_e(i), t(i+1)]}$ is found from the drawing in
    Fig.~\ref{fig:commutationex1.1d.graph}, where $P_e(i)$ and $t(i+1)$
    are represented with arrows.
    We treat the arrows as unit vectors, and find the sum of scalar products of unit vectors defined
    at the same sites, which after being multiplied by ${\hbar}/(2^2i)$, lead to the commutation relation.
    From the graphical rule and the symmetry of the modes, the following commutations are obtained,
    where $a, b \in \{e, t\}$.
    \begin{eqnarray}
    {[P_{a}(i), b(j)]}&=& {[P_{b}(j), a(i)]},  \\
    {[P_e(i), t(j)]} &=& -{[P_{t}(i), e(j)]}. \label{eq:followequation1}
    \end{eqnarray}
The commutation relations in reciprocal space are calculated from
the relations,
\begin{equation}
[P_a(k),b(k')]=\delta_{k',-k}\sum_j [P_a(i=0),b(j)]e^{ikj},
\end{equation}
which are shown in Table~\ref{table:commutation.1d}.
The results found here are applicable, for example, for the study of quantum mechanical dynamics
of non-linear excitations mentioned in Section II.A.

  \renewcommand{\arraystretch}{1.2}
  \renewcommand{\baselinestretch}{1.6}
  \begin{table}[htb]
    \centering
  \begin{tabular}{c|c c}\hline \hline
      & {$ \frac{i}{\hbar} P_e(k)$} & {$ \frac{i}{\hbar} P_{t}(k)$} \\
    \hline
    $e(-k)$ & $\frac{1}{2} (1-\cos{k})$ & $-\frac{i}{2} \sin{k}  $  \\
    $t(-k)$ & $ \frac{i}{2} \sin{k} $ &  $\frac{1}{2} (1+\cos{k}) $  \\ \hline \hline
  \end{tabular}
    \caption{Commutation relation, $[\frac{i}{\hbar}P_a(k),b(-k)]$, between symmetry modes and their
    conjugate momenta for the one-dimensional chain in reciprocal space.}
    \label{table:commutation.1d}
  \end{table}

\subsection{Two-dimensional square lattice with a monatomic basis}

Quantum mechanical nature of lattice is also important for two or three-dimensional
lattices, for example, near the structural phase transitions.
In this subsection, we find quantum mechanical commutation relations
for the symmetry modes and their conjugate momenta
for the square lattice studied in Section II.B.

Conjugate momenta for the atomic scale symmetry modes are as follows.
    \begin{eqnarray*}
        P_{e_1}(\vec{i})=\frac{M}{4} \dot{e}_1(\vec{i}) &=& \frac{1}{8\sqrt{2}}\left(-p^{x}_{\vec{i}}-p^{y}_{\vec{i}}+p^{x}_{\vec{i}+10}-p^{y}_{\vec{i}+10} \right.\\
            &&\left.-p^{x}_{\vec{i}+01}+p^{y}_{\vec{i}+01}+p^{x}_{\vec{i}+11}+p^{y}_{\vec{i}+11}\right), \\
        P_{e_2}(\vec{i}) =\frac{M}{4} \dot{e}_2(\vec{i})&=& \frac{1}{8\sqrt{2}}\left(-p^{x}_{\vec{i}}-p^{y}_{\vec{i}}-p^{x}_{\vec{i}+10}+p^{y}_{\vec{i}+10} \right. \\
            &&\left. +p^{x}_{\vec{i}+01}-p^{y}_{\vec{i}+01}+p^{x}_{\vec{i}+11}+p^{y}_{\vec{i}+11}\right), \\
        P_{e_3}(\vec{i}) =\frac{M}{4} \dot{e}_3(\vec{i})&=& \frac{1}{8\sqrt{2}}\left(-p^{x}_{\vec{i}}+p^{y}_{\vec{i}}+p^{x}_{\vec{i}+10}+p^{y}_{\vec{i}+10} \right. \\
            &&\left. -p^{x}_{\vec{i}+01}-p^{y}_{\vec{i}+01}+p^{x}_{\vec{i}+11}-p^{y}_{\vec{i}+11}\right), \\
        P_{r  }(\vec{i})=\frac{M}{4} \dot{r}(\vec{i}) &=& \frac{1}{8\sqrt{2}}\left( p^{x}_{\vec{i}}-p^{y}_{\vec{i}}+p^{x}_{\vec{i}+10}+p^{y}_{\vec{i}+10} \right. \\
            && \left. -p^{x}_{\vec{i}+01}-p^{y}_{\vec{i}+01}-p^{x}_{\vec{i}+11}+p^{y}_{\vec{i}+11}\right), \\
        P_{s_x}(\vec{i}) =\frac{M}{4} \dot{s}_x(\vec{i})&=& \frac{1}{8}\left(p^{x}_{\vec{i}}-p^{x}_{\vec{i}+10} -p^{x}_{\vec{i}+01}+p^{x}_{\vec{i}+11}\right), \\
        P_{s_y}(\vec{i})=\frac{M}{4} \dot{s}_y(\vec{i}) &=& \frac{1}{8}\left(p^{y}_{\vec{i}}-p^{y}_{\vec{i}+10} -p^{y}_{\vec{i}+01}+p^{y}_{\vec{i}+11}\right), \\
        P_{t_x}(\vec{i})=\frac{M}{4} \dot{t}_x(\vec{i}) &=& \frac{1}{8}\left( p^{x}_{\vec{i}}+p^{x}_{\vec{i}+10}+p^{x}_{\vec{i}+01}+p^{x}_{\vec{i}+11}\right), \\
        P_{t_y}(\vec{i})=\frac{M}{4} \dot{t}_y(\vec{i}) &=& \frac{1}{8}\left(p^{y}_{\vec{i}}+p^{y}_{\vec{i}+10}+p^{y}_{\vec{i}+01}+p^{y}_{\vec{i}+11}\right).
    \end{eqnarray*}
From the fundamental commutation relations for displacement operators and momentum operators,
    \begin{eqnarray*}
        {[p_{\vec{i}}^x, u^x_{\vec{j}}]} = [p_{\vec{i}}^y, u^y_{\vec{j}}] &=& \frac{\hbar}{i} \delta_{\vec{i},\vec{j}}, \\
        {[p_{\vec{i}}^x, u^y_{\vec{j}}]} = [p_{\vec{i}}^y, u^x_{\vec{j}}] &=& 0,
    \end{eqnarray*}
the commutation relations between modes and their conjugate momenta
are calculated in a straight forward way.

\begin{figure}
        \includegraphics[scale=1.0, clip=true]{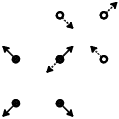}
        \caption{Graph to find commutation relation, $[P_{e_1}(\vec{i}),e_2(\vec{i}+11)]$.}
        \label{fig:commutationex1.2d.graph}
    \end{figure}

However, it is more convenient to use the graphical method, explained for
the one-dimensional chain in the previous subsection.
The above fundamental commutation relations for $\vec{i}=\vec{j}$ have the form of
    \begin{eqnarray*}
        \hat{x}\cdot\hat{x}=\hat{y}\cdot\hat{y}&=&1, \\
        \hat{x}\cdot\hat{y}=\hat{y}\cdot\hat{x}&=&0,
    \end{eqnarray*}
    except for the factor ${\hbar}/{i}$, where $\hat{x}$ and $\hat{y}$
    represent unit vectors, not operators.
    Therefore, the commutation relation $[P_a(\vec{i}), b(\vec{j})]$,
    where $a$ and $b$ represent the eight atomic scale
    modes, is found from the drawings of $a$ and $b$ modes on the
    square lattice. The sum of the scalar products of the unit vectors
    at the sites shared
    by the two modes, multiplied by $\hbar/(4^2 i)$, gives the commutation of the
    two operators.
    (The multiplication factor after ${\hbar}/i$
    is associated with the number of atoms in the motif for the lattice with a monatomic basis,
    that is 4 for the square lattice and 2 for the chain.)
    For example,
    $[P_{e1}(\vec{i}), e_2(\vec{i}+11)]$ is found from Fig.~\ref{fig:commutationex1.2d.graph}, as follows
   \begin{equation}
        {[P_{e1}(\vec{i}), e_2(\vec{i}+11)]} =  \frac{\hbar}{4^2i} (-1).
    \end{equation}

  \renewcommand{\arraystretch}{1.2}
  \renewcommand{\baselinestretch}{1.6}
  \begin{table*}[ht]
    \centering
  \begin{tabular}{c|c c c c | c c c c}\hline \hline
      & {$ \frac{i}{\hbar} P_{e1}(\vec{k})$}
      & {$ \frac{i}{\hbar} P_{e2}(\vec{k}) $}
      & {$ \frac{i}{\hbar} P_{e3}(\vec{k}) $}
      & {$ \frac{i}{\hbar} P_{r}(\vec{k}) $}

      & {$ \frac{i}{\hbar} P_{sx}(\vec{k})$}
      & {$ \frac{i}{\hbar} P_{sy}(\vec{k}) $}
      & {$ \frac{i}{\hbar} P_{tx}(\vec{k}) $}
      & {$ \frac{i}{\hbar} P_{ty}(\vec{k}) $}
      \\
    \hline

    $e_1(-\vec{k})$
    & $\frac{1 - C_{kx} C_{ky}}{4}$
    & $\frac{S_{kx} S_{ky}}{4}$
    & $\frac{- C_{kx} + C_{ky}}{4}$
    & 0

    & $\frac{i (1-C_{kx}) S_{ky}} {4\sqrt{2}}$
    & $\frac{i (1-C_{ky}) S_{kx}} {4\sqrt{2}}$
    & $\frac{i (1+C_{ky}) S_{kx}} {-4\sqrt{2}}$
    & $\frac{i (1+C_{kx}) S_{ky}} {-4\sqrt{2}}$
    \\

    $e_2(-\vec{k})$
    & $\frac{S_{kx}S_{ky}}{4}$
    & $\frac{1-C_{kx}C_{ky}}{4}$
    & 0
    & $\frac{-C_{kx}+C_{ky}}{4}$

    & $\frac{i (1-C_{ky}) S_{kx}} {4\sqrt{2}}$
    & $\frac{i (1-C_{kx}) S_{ky}} {4\sqrt{2}}$
    & $\frac{i (1+C_{kx}) S_{ky}} {-4\sqrt{2}}$
    & $\frac{i (1+C_{ky}) S_{kx}} {-4\sqrt{2}}$
    \\

    $e_3(-\vec{k})$
    & $\frac{-C_{kx}+C_{ky}}{4}$
    & 0
    & $\frac{1-C_{kx}C_{ky}}{4}$
    & $\frac{-S_{kx}S_{ky}}{4}$

    & $\frac{i (1-C_{kx}) S_{ky}} {4\sqrt{2}}$
    & $\frac{i (1-C_{ky}) S_{kx}} {-4\sqrt{2}}$
    & $\frac{i (1+C_{ky}) S_{kx}} {-4\sqrt{2}}$
    & $\frac{i (1+C_{kx}) S_{ky}} {4\sqrt{2}}$
    \\

    $r(-\vec{k})$
    & 0
    & $\frac{-C_{kx} + C_{ky}}{4}$
    & $\frac{-S_{kx}S_{ky}}{4}$
    & $\frac{1-C_{kx}C_{ky}}{4}$

    & $\frac{i (1-C_{ky}) S_{kx}} {-4\sqrt{2}}$
    & $\frac{i (1-C_{kx}) S_{ky}} {4\sqrt{2}}$
    & $\frac{i (1+C_{kx}) S_{ky}} {4\sqrt{2}}$
    & $\frac{i (1+C_{ky}) S_{kx}} {-4\sqrt{2}}$
    \\

    \hline

    $s_x(-\vec{k})$
    & $\frac{i (1-C_{kx}) S_{ky}} {-4\sqrt{2}}$
    & $\frac{i (1-C_{ky}) S_{kx}} {-4\sqrt{2}}$
    & $\frac{i (1-C_{kx}) S_{ky}} {-4\sqrt{2}}$
    & $\frac{i (1-C_{ky}) S_{kx}} {4\sqrt{2}}$

    & $\frac{(1-C_{kx})(1-C_{ky})}{4}$
    & 0
    & $\frac{-S_{kx}S_{ky}}{4}$
    & 0

    \\

    $s_y(-\vec{k})$
    & $\frac{i (1-C_{ky}) S_{kx}} {-4\sqrt{2}}$
    & $\frac{i (1-C_{kx}) S_{ky}} {-4\sqrt{2}}$
    & $\frac{i (1-C_{ky}) S_{kx}} {4\sqrt{2}}$
    & $\frac{i (1-C_{kx}) S_{ky}} {-4\sqrt{2}}$

    & 0
    & $\frac{(1-C_{kx})(1-C_{ky})}{4}$
    & 0
    & $\frac{-S_{kx}S_{ky}}{4}$

    \\

    $t_x(-\vec{k})$
    & $\frac{i (1+C_{ky}) S_{kx}} {4\sqrt{2}}$
    & $\frac{i (1+C_{kx}) S_{ky}} {4\sqrt{2}}$
    & $\frac{i (1+C_{ky}) S_{kx}} {4\sqrt{2}}$
    & $\frac{i (1+C_{kx}) S_{ky}} {-4\sqrt{2}}$

    & $\frac{-S_{kx}S_{ky}}{4}$
    & 0
    & $\frac{(1+C_{kx})(1+C_{ky})}{4}$
    & 0

    \\

    $t_y(-\vec{k})$
    & $\frac{i (1+C_{kx}) S_{ky}} {4\sqrt{2}}$
    & $\frac{i (1+C_{ky}) S_{kx}} {4\sqrt{2}}$
    & $\frac{i (1+C_{kx}) S_{ky}} {-4\sqrt{2}}$
    & $\frac{i (1+C_{ky}) S_{kx}} {4\sqrt{2}}$

    & 0
    & $\frac{-S_{kx}S_{ky}}{4}$
    & 0
    & $\frac{(1+C_{kx})(1+C_{ky})}{4}$

    \\
    \hline \hline
    \end{tabular}
    \caption{Commutation relation, $[\frac{i}{\hbar} P_a(\vec{k}), b(-\vec{k})]$, between symmetry modes and their conjugate momenta
    for the two-dimensional square lattice in reciprocal space.
    $C_{kx}$, $C_{ky}$, $S_{kx}$, and $S_{ky}$ represent $\cos k_x$, $\cos k_y$, $\sin k_x$, and $\sin k_y$, respectively.}
    \label{table:commutation.2d}
  \end{table*}

   Presented graphical method is also useful to find the following symmetry-related properties of the commutation relations,
   where $a$ and $b$ represent any of the eight modes, and
   \textit{even} and \textit{odd} represent the modes with even symmetry under point reflection, namely, $e_1$, $e_2$, $e_3$, $r$,
    and the modes with odd symmetry, namely, $s_x$, $s_y$, $t_x$, $t_y$, respectively.
       \begin{eqnarray*}
        {[P_a(\vec{i}), b(\vec{i})]}&=&\frac{1}{4}\frac{\hbar}{i}\delta_{ab}, \\
        {[P_a(\vec{i}), b(\vec{j})]}&=&{[P_b(\vec{j}), a(\vec{i})]}, \\
        {[P_{even}(\vec{i}), even'(\vec{j})]}&=&{[P_{even'}(\vec{i}), even(\vec{j})]}, \\
        {[P_{even}(\vec{i}), odd(\vec{j})]}&=&-{[P_{odd}(\vec{i}), even(\vec{j})]}, \\
        {[P_{odd}(\vec{i}), odd'(\vec{j})]}&=&{[P_{odd'}(\vec{i}), odd(\vec{j})]}.
    \end{eqnarray*}
    The commutation relations in reciprocal space are found from the relation
  \begin{equation}
  [P_a(\vec{k}),b(\vec{k'})]=\delta_{\vec{k'},-\vec{k}} \sum_{\vec{j}} [P_a(\vec{i}=0),b(\vec{j})]e^{i\vec{k}\cdot \vec{j}},
  \end{equation}
   which are provided in Table~\ref{table:commutation.2d}.

\section{SUMMARY}
    In this article, we have presented mode-based atomic scale description of
    lattice dynamics. It is found that not only the potential energy but also
    the kinetic energy is described in terms of the atomic scale modes, for
    which the inclusion of the rigid modes is essential.
    This approach has been
    demonstrated
    for the one-dimensional chain and the two-dimensional
    square lattice with a monatomic basis.
    The comparison with a continuum model has shown that our approach is suitable
    for multiscale description of lattice dynamics.
    The approach has been extended to quantum mechanics, and the
    commutation relations have been obtained.
    We expect that this method would be useful in describing
    atomic scale lattice dynamics in systems with strong anharmonicity and complex energy landscape,
    which can be compared with the results from experiments, such as, time-resolved x-ray diffraction.


\appendix

\section{Example of the application of symmetry modes for lattice dynamics - Phonon mode analysis}
As a simple demonstration for the application of symmetry modes,
we analyze phonon modes in terms of atomic scale symmetry modes
for the square lattice with a harmonic potential shown below.~\cite{Ahn03}
    \begin{eqnarray}
        V_{\rm sq} = \sum_{\vec {i}}&& \frac{1}{2} A_{1} e_{1}(\vec {i})^{2} +
        \frac{1}{2} A_{2} e_{2}(\vec{i})^{2} + \frac{1}{2} A_{3} e_{3}(\vec{i})^{2} \nonumber\\
        &&+ \frac{1}{2} B [ s_{x}(\vec{i})^{2} + s_{y}(\vec{i})^{2} ].
        \label{eq:sq.lat.potential}
    \end{eqnarray}
The phonon dispersion relation for this potential energy was presented
in Ref.~\onlinecite{Ahn03}.
Furthermore, the phonon mode at
$\vec{k}=(\pi,\pi)$, both in the upper and the lower branch, was shown entirely
composed of short wave length modes, $s_x$ and $s_y$, due to the constraints.
In general, at other $\vec{k}$ points,
the contribution of different symmetry modes to the phonon mode depends not only
on the constraint equations, but also on the values of the elastic moduli in the energy expression.
In this Appendix, we present our current study on such general expressions which 
describes how different symmetry modes contribute to phonon modes for the entire first Brillouin zone
for the potential $V_{\rm sq}$ shown above.

\begin{figure}[b]
        \includegraphics[width=3.4in, clip=true]{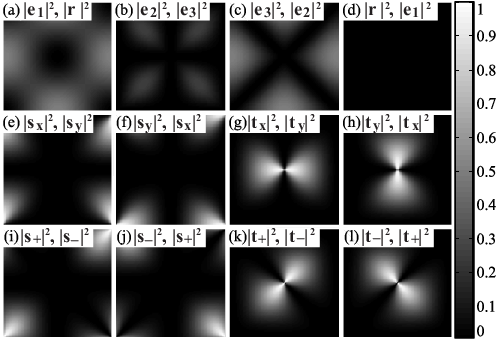}
        \caption{Normalized symmetry-mode squared amplitude for phonons within the first Brillouin zone of the square lattice
        for $A_2=A_3$, that is, $a=1$, where $a=(A_1-A_2+A_3)/(A_1+A_2-A_3)$.
        The center of each square corresponds to $\vec{k}=(0,0)$ and the four corners $\vec{k}=(\pm\pi,\pm\pi)$.
        Different modes in the upper branch and the lower branch have the same amplitude.
        The first and the second label for each panel indicate the mode for upper branch and lower branch, respectively.
        Therefore, for the upper branch, the twelve panels correspond to
        (a)$|e_1|^2$, (b)$|e_2|^2$, (c)$|e_3|^2$, (d)$|r|^2$,
        (e)$|s_x|^2$, (f)$|s_y|^2$, (g)$|t_x|^2$, (h)$|t_y|^2$,
        (i)$|s_+|^2$, (j)$|s_-|^2$, (k)$|t_+|^2$, and (l)$|t_-|^2$.
        For the lower branch, they correspond to
        (a)$|r|^2$, (b)$|e_3|^2$, (c)$|e_2|^2$, (d)$|e_1|^2$,
        (e)$|s_y|^2$, (f)$|s_x|^2$, (g)$|t_y|^2$, (h)$|t_x|^2$,
        (i)$|s_-|^2$, (j)$|s_+|^2$, (k)$|t_-|^2$, and (l)$|t_+|^2$.
        Panel (d) shows that the upper and lower branches of phonon modes remain locally rotationless and area-preserving, respectively, in this particular case of $A_2=A_3$.
        This feature can be understood easily
        in the long wavelength limit
        from the fact that the sound velocity is isotropic for
        $A_2=A_3$. See Footnote~(\onlinecite{isotropic}) for details.
        Such isotropic sound velocity for crystals can be useful for device applications.
        }
        \label{fig:modes-a1-up}
    \end{figure}

The equation of motion using conventional displacement-based approach
leads the expression for the normal mode at $\vec{k}$,
$[u_x^\pm(\vec{k}),u_x^\pm(\vec{k})]$, where $\pm$ represent the upper and lower branch,
except for an overall factor.
By using the $\vec{k}$-space relation between symmetry modes and displacements,
which is obtained from Eqs.~(\ref{eq:def-e1})-(\ref{eq:def-r}),
we get the expression of
$[e_1^\pm(\vec{k}),e_2^\pm(\vec{k}),e_3^\pm(\vec{k}),r^\pm(\vec{k}),
s_x^\pm(\vec{k}),s_y^\pm(\vec{k}),t_x^\pm(\vec{k}),t_y^\pm(\vec{k})]$
for the phonon modes at $\vec{k}$, where $\pm$ again represents the upper and the lower branches.~\cite{mode-amplitude}
The results are shown in Table~\ref{table:modeamp},
where $\beta_1=1-\cos k_x \cos k_y$, $\beta_2=-\sin k_x \sin k_y$,
$\beta_3=\cos k_x - \cos k_y$, $\beta_4=(1-\cos k_x)(1-\cos k_y)$,
$\beta_5=(1+\cos k_x)(1+\cos k_y)$, and $a=(A_1-A_2+A_3)/(A_1+A_2-A_3)$.
In this result, the overall factor of the normal mode is determined
by the normalization condition $|u_x^\pm(\vec{k})|^2+|u_x^\pm(\vec{k})|^2=4$,
that is, $|e_1^\pm(\vec{k})|^2+|e_2^\pm(\vec{k})|^2+|e_3^\pm(\vec{k})|^2+|r^\pm(\vec{k})|^2
+|s_x^\pm(\vec{k})|^2+|s_y^\pm(\vec{k})|^2+|t_x^\pm(\vec{k})|^2+|t_y^\pm(\vec{k})|^2=1$.
To be specific,
we show these expressions for a special case of $a=1$,
which corresponds to $A_2=A_3$,
in the last column of Table~\ref{table:modeamp}
and plot them within the first Brillouin zone of the square lattice
in Fig.~\ref{fig:modes-a1-up}.
The results show the contributions of different symmetry modes to the upper and the lower branch
phonon modes within the first Brillouin zone.
Some of the features are discussed in the figure caption and Footnote~\onlinecite{isotropic}.
 \renewcommand{\arraystretch}{1.2}
 \renewcommand{\baselinestretch}{1.6}
  \begin{table*}[htb]
    \centering
  \begin{tabular}
  {c|c|c c}
  \hline \hline
  Mode
  & General expression
  & \multicolumn{2}{c}{Special case: $a=1$}
  \\
       & upper/lower branch
       & upper branch
       & lower branch
       \\
  \hline
$|e_1(\vec{k})|^2$
& $\frac{1}{8}\left( \beta_1 \pm \frac{\beta_2^2 + a \beta_3^2}{\sqrt{\beta_2^2 + a^2 \beta_3^2}}\right) $
& $\frac{\beta_1}{4}$
& 0
\\
$|e_2(\vec{k})|^2$
& $\frac{1}{8}\left( \beta_1 \pm \frac{\beta_2^2 - a \beta_3^2}{\sqrt{\beta_2^2 + a^2 \beta_3^2}}\right) $
& $\frac{\beta_2^2}{4\beta_1}$
& $\frac{\beta_3^2}{4\beta_1}$
\\
$|e_3(\vec{k})|^2$
& $\frac{1}{8}\left( \beta_1 \mp \frac{\beta_2^2 - a \beta_3^2}{\sqrt{\beta_2^2 + a^2 \beta_3^2}}\right) $
& $\frac{\beta_3^2}{4\beta_1}$
& $\frac{\beta_2^2}{4\beta_1}$
\\
$|r(\vec{k})|^2$
& $\frac{1}{8}\left( \beta_1 \mp \frac{\beta_2^2 + a \beta_3^2}{\sqrt{\beta_2^2 + a^2 \beta_3^2}}\right) $
& 0
& $\frac{\beta_1}{4}$
\\
\hline
$|s_x(\vec{k})|^2$
& $\frac{\beta_4}{8}\left( 1 \mp \frac{ a \beta_3}{\sqrt{\beta_2^2 + a^2 \beta_3^2}}\right) $
& $\frac{\beta_4}{8\beta_1}\left(\beta_1-\beta_3 \right)$
& $\quad\frac{\beta_4}{8\beta_1}\left(\beta_1+\beta_3 \right)$
\\
$|s_y(\vec{k})|^2$
& $\frac{\beta_4}{8}\left( 1 \pm \frac{ a \beta_3}{\sqrt{\beta_2^2 + a^2 \beta_3^2}}\right) $
& $\frac{\beta_4}{8\beta_1}\left(\beta_1+\beta_3 \right)$
& $\quad\frac{\beta_4}{8\beta_1}\left(\beta_1-\beta_3 \right)$
\\
$|t_x(\vec{k})|^2$
& $\frac{\beta_5}{8}\left( 1 \mp \frac{ a \beta_3}{\sqrt{\beta_2^2 + a^2 \beta_3^2}}\right) $
& $\frac{\beta_5}{8\beta_1}\left(\beta_1-\beta_3 \right)$
& $\quad\frac{\beta_5}{8\beta_1}\left(\beta_1+\beta_3 \right)$
\\
$|t_y(\vec{k})|^2$
& $\frac{\beta_5}{8}\left( 1 \pm \frac{ a \beta_3}{\sqrt{\beta_2^2 + a^2 \beta_3^2}}\right) $
& $\frac{\beta_5}{8\beta_1}\left(\beta_1+\beta_3 \right)$
& $\quad\frac{\beta_5}{8\beta_1}\left(\beta_1-\beta_3 \right)$
\\
\hline
$|s_+(\vec{k})|^2$
& $\frac{\beta_4}{8}\left( 1 \mp \frac{ \beta_2}{\sqrt{\beta_2^2 + a^2 \beta_3^2}}\right) $
& $\frac{\beta_4}{8\beta_1}\left(\beta_1-\beta_2 \right)$
& $\quad\frac{\beta_4}{8\beta_1}\left(\beta_1+\beta_2 \right)$
\\
$|s_-(\vec{k})|^2$
& $\frac{\beta_4}{8}\left( 1 \pm \frac{ \beta_2}{\sqrt{\beta_2^2 + a^2 \beta_3^2}}\right) $
& $\frac{\beta_4}{8\beta_1}\left(\beta_1+\beta_2 \right)$
& $\quad\frac{\beta_4}{8\beta_1}\left(\beta_1-\beta_2 \right)$
\\
$|t_+(\vec{k})|^2$
& $\frac{\beta_5}{8}\left( 1 \mp \frac{ \beta_2}{\sqrt{\beta_2^2 + a^2 \beta_3^2}}\right) $
& $\frac{\beta_5}{8\beta_1}\left(\beta_1-\beta_2 \right)$
& $\quad\frac{\beta_5}{8\beta_1}\left(\beta_1+\beta_2 \right)$
\\
$|t_-(\vec{k})|^2$
& $\frac{\beta_5}{8}\left( 1 \pm \frac{ \beta_2}{\sqrt{\beta_2^2 + a^2 \beta_3^2}}\right) $
& $\frac{\beta_5}{8\beta_1}\left(\beta_1+\beta_2 \right)$
& $\quad\frac{\beta_5}{8\beta_1}\left(\beta_1-\beta_2 \right)$
\\
\hline
  \end{tabular}
    \caption{
    Normalized symmetry-mode squared amplitude for phonons within the first Brillouin zone of the square lattice.
    For $\pm$ and $\mp$, the upper sign corresponds to the upper branch, and the lower sign the lower branch.
    $\beta_1=1-\cos k_x \cos k_y$,
$\beta_2=-\sin k_x \sin k_y$, $\beta_3=\cos k_x - \cos k_y$, $\beta_4=(1-\cos k_x)(1-\cos k_y)$,
$\beta_5=(1+\cos k_x)(1+\cos k_y)$, and
$a=(A_1-A_2+A_3)/(A_1+A_2-A_3)$. Special case of $a=1$ corresponds to the case of $A_2=A_3$.
}
    \label{table:modeamp}
  \end{table*}

\begin{thebibliography}{99}
\bibitem{Rini07}
M. Rini, R. Tobey, N. Dean, J. Itatani, Y. Tomioka, Y. Tokura, R. W. Schoenlein, and A. Cavalleri,
Nature (London) 449, 72 (2007).
    \bibitem{Salamon01}
    M. B. Salamon and M. Jaime, Rev. Mod. Phys. {\bf 73}, 583 (2001).
    \bibitem{Millis98}
    A. J. Millis, Nature (London) {\bf 392}, 147 (1998).
    \bibitem{Jin94}
    S. Jin, T. H. Tiefel, M. McCormack, R. A. Fastnacht, R. Ramesh, and L. H. Chen,
    Science {\bf 264}, 413 (1994).
    \bibitem{Tranquada95}
    J. M. Tranquada, B. J. Sternlieb, J. D. Axe, Y. Nakamura, and S. Uchida,
    Nature (London) {\bf 375}, 561 (1995).
    \bibitem{Kivelson03}
    S. A. Kivelson, I. P. Bindloss, E. Fradkin, V. Oganesyan, J. M. Tranquada, A. Kapitulnik, and C. Howald,
    Rev. Mod. Phys. {\bf 75}, 1201 (2003).
    \bibitem{Kiryukhin04}
    V. Kiryukhin, New J. Phys. {\bf 6} 155 (2004).
    \bibitem{Ahn04}
    K. H. Ahn, T. Lookman and A. R. Bishop, Nature (London) {\bf 428},  401 (2004).
       \bibitem{Gaffney05}
    See, e.g., K. J. Gaffney and H. N. Chapman, Science {\bf 316}, 1444 (2005), and references therein.
    \bibitem{Shenoy99}
    S.R. Shenoy, T. Lookman, A. Saxena, and A. R. Bishop, Phys. Rev. B {\bf 60}, R12 537  (1999).
    \bibitem{Lookman03}
     T. Lookman, S. R. Shenoy, K. {\O}. Rasmussen, A. Saxena, and A. R. Bishop, Phys. Rev. B {\bf 67}, 024114 (2003).
    \bibitem{Ahn03}
        K. H. Ahn, T. Lookman, A. Saxena, and A. R. Bishop,
        Phys. Rev. B {\bf 68}, 092101 (2003).
    \bibitem{Ahn05}
    K. H. Ahn, T. Lookman, A. Saxena and A. R. Bishop, Phys. Rev. B {\bf 71}, 212102 (2005).
    \bibitem{Zhu03}
    J.-X. Zhu, K. H. Ahn, Z. Nussinov, T. Lookman, A. V. Balatsky and A. R. Bishop,
    Phys. Rev. Lett. {\bf 91}, 057004 (2003).
    \bibitem{Doh07}
    H. Doh, Y. B. Kim, and K. H. Ahn, Phys. Rev. Lett. {\bf 98}, 126407 (2007).
    \bibitem{Kittel}
        C. Kittel, {\it Introduction to Solid State Physics}, 8th ed. (John Wiley and Sons, Inc., Singapore, 1986).
    \bibitem{Chen96} D. Chen, S. Aubry, and G. P. Tsironis, Phys. Rev. Lett. {\bf 77}, 4776 (1996).
    \bibitem{Kosevich08}  Y. A. Kosevich, L. I. Manevitch, and A. V. Savin, Phys. Rev. E {\bf 77}, 046603 (2008).
    \bibitem{footnote}
        We note that inverting the relation between
$(s_x(\vec{k}), s_y(\vec{k}))$
and $(u^x_{\vec{k}}, u^y_{\vec{k}})$ is not possible for certain wave vectors,
for example, wave vectors with $k_x$=0 or $k_y$=0.
In those cases, new constraint equations should be found from the definition
of the modes.
\bibitem{mode-amplitude}
This can be also done without explicit use of displacement variables,
as suggested by Eq.~(\ref{eq:Lagrangian}) and done for the 1-D chain.
\bibitem{isotropic}
   If the two shape-changing modes, $e_2$ and $e_3$, have identical moduli,
the lattice sustains isotropic phonon dispersion in the long wavelength limit,
in which the lattice behaves like an isotropic continuous medium.
Such medium would support longitudinal phonon modes in the upper branch and
transverse phonon modes in the lower branch:
the former rotationless and the latter locally area-preserving.
In finite wavelength limit case, the dispersion relation of phonon modes
become anisotropic, reflecting only 90$^o$ rotational symmetry of the square lattice,
but the absence of rotational and area changing mode
for the upper and the lower branch phonons, respectively, still holds.
\end{thebibliography}
\end{document}